\documentclass[aps,pra,preprint,floatfix]{revtex4}
\usepackage{graphicx,amssymb,amsmath}
\usepackage{subfigure}

\newcommand{\I}{\mathrm i}
\newcommand {\pt}{\partial}

\newcommand{\la}{\lambda}
\newcommand{\mr}{\mathrm{R}}
\newcommand{\ml}{\mathrm{L}}

\begin{document}

\title{{\it Preprint}\\Right and left inverse scattering problems formulations for the Zakharov--Shabat system}

\author{A.\,E.~Chernyavsky}
\affiliation{Institute of Automation and Electrometry, 
Siberian Branch, Russian Academy of Sciences,
1 Koptjug Avenue, Novosibirsk 630090, Russia}

\author{L.\,L.~Frumin}
\affiliation{Institute of Automation and Electrometry, 
	Siberian Branch, Russian Academy of Sciences,
	1 Koptjug Avenue, Novosibirsk 630090, Russia}

\author{A.\,V.~Gelash}
\affiliation{Institute of Automation and Electrometry, 
Siberian Branch, Russian Academy of Sciences,
1 Koptjug Avenue, Novosibirsk 630090, Russia}

\begin{abstract}
	We consider right and left formulations of the inverse scattering problem for the Zakharov--Shabat system and the corresponding integral Gelfand--Levitan--Marchenko equations. Both formulations are helpful for numerical solving of the inverse scattering problem, which we perform using the previously developed Toeplitz Inner Bordering (TIB) algorithm. First, we establish general relations between the right and left scattering coefficients. Here, along with the known results, we introduce a relation between the left and right norming coefficients for the N -soliton solution.Then we propose an auxiliary kernel of the left Gelfand--Levitan--Marchenko equations, which allows one to solve the right scattering problem numerically. We generalize the TIB algorithm, initially proposed in the left formulation, to the right scattering problem case with the obtained formulas. The test runs of the TIB algorithm illustrate our results reconstructing the various nonsymmetrical potentials from their right scattering data.
\end{abstract}
\maketitle


\section{Introduction} \label{s:1}
The direct and inverse scattering problems (SPs) form the basis of the Inverse Scattering Transform (IST) method -- an outstanding achievement of modern mathematical physics, which allows integrating (solving) certain nonlinear partial differential equations. 
Examples of such ``integrable'' equations represent well-known models of nonlinear mathematical physics: the Korteweg-De Vries equation, the sine-Gordon equation, and the nonlinear Schroedinger equation (NLSE), see \cite{Zakharov:1984,Lamb:1980,Ablowitz:1981,Zakharov:1971}. 
From a mathematical point of view, the IST represents direct, and inverse SPs formulated for the systems of linear partial differential equations and the integral Gelfand--Levitan--Marchenko equations. The latter being Fredholm's integral equation of the second type poses though well-posed but challenging inverse problems from a numerical point of view.

In this work, we consider the IST for the focusing and defocusing versions of the NLSE, which we write in the following form
\begin{equation}\label{NLSE}
\I\frac{\pt u}{\pt t} +\frac{\pt^2 u}{\pt x^2} - 2\sigma u\,|u|^2=0,
\end{equation}
where $u=u(x,t)$ is complex wave field, $\I$ is the imaginary unit, $x$ and $t$ are space and time coordinates.
The constant $\sigma$ is equal to $+1$ or $-1$, respectively, in the cases of the defocusing and the focusing NLSE.

The IST for the NLSE (\ref{NLSE}) is formulated using the Zakharov and Shabat (ZS) system of linear partial differential equations \cite{Zakharov:1971}, which at a fixed moment of time reads as,
\begin{equation}\label{ZS}
\Phi_x=\widehat{\mathbf{U}} \Phi, \qquad
\widehat{\mathbf{U}}=
\left[
\begin{array}{cc}
-\I\la & u \\
\sigma u^* & \I\la  \end{array}
 \right].
\end{equation}
Here $\Phi(x) = (\phi_1(x),\phi_2(x))^{\mathrm{T}}$ (the superscript $\mathrm{T}$ means vector transposing) is a two-component spectral wave function, $\la=\xi+\I\eta$ is a complex spectral parameter and the wave field $u=u(x)$ plays the role of a potential. The ZS system establishes direct and inverse SPs for the potential and its spectral (scattering) data, which can be resolved analytically only in a few specific cases: solitonic, rectangular and hyperbolic secant potentials \cite{Manakov:1974,Satsuma:1974,Zakharov:1984}.

Numerical approaches to the SPs provide an efficient tool for solving the Cauchy problem without time iterations \cite{Trogdon:2013} and
also for analysis and synthesis of nonlinear waves \cite{Braud:2016,Turitsyn:2017,Gelash:2018,Suret:2020,Slunyaev:2021}.
In addition, some applied physical issues can be studied using these approaches since the ZS system (\ref{ZS}) coincides with the Kogelnik equations for the coupled-mode model \cite{Kogelnik:1969},  which describe the scattering of waves on Bragg gratings \cite{Gorbenko:2019,Kashyap:1999}.

The inverse and direct SPs for the ZS system can be solved numerically using the Toeplitz inner bordering (TIB) algorithms \cite{Belay:2007,Frumin:2015}, see also supplementary material in \cite{Turitsyn:2017}. The TIB algorithm effectively employs the Toeplitz symmetry of discretized systems of Gelfand--Levitan--Marchenko integral equations. It belongs to the family of fast algorithms of the Levinson type \cite{Levinson:1947,Blahut:1985}. It requires only $O(N^2)$ arithmetic operations to solve the inverse SP, where $N$ is the number of discrete points of the problem, providing the second-order computation accuracy. Moreover, the inversion of the steps of the inverse problem algorithm provides the solution to the direct SP \cite{Frumin:2015}. Today, the TIB algorithms are used in various applied problems, including the synthesis of Bragg gratings \cite{Belay:2007,Buryak:2009,Belay:2010}, the fast solution of the inverse problem for the Helmholtz equations \cite{Belay:2008}, the development of new nonlinear approaches for efficient information transmission in fiber-optic communication lines \cite{Turitsyn:2017,Aref:2018,Frumin:2017,Le:2014,Zhang:2020,Bogdanov:2021,Delitsyn:2022}. Recently, the TIB algorithms have been generalized to the case of the vector NLS equation (the Manakov system), which describes polarized nonlinear waves \cite{Frumin:2021}.

We do not consider here other algorithms, such as "superfast layer peeling" (which, although faster than TIB, is less stable. A brief but informative overview of this and other approaches and algorithms for solving the inverse scattering problem for the Zakharov--Shabat system is given in a recent review by Delitsyn \cite{Delitsyn:2022}.

The TIB algorithm was initially proposed in the left formulation of the SP, which is more convenient for it; see details in \cite{Belay:2007,Frumin:2015}. However, many studies use another -- right formulation of the SP. In particular, theoretical works usually present the right scattering coefficients, e.g., \cite{Manakov:1974,Satsuma:1974}. To fill this gap and extend the TIB algorithm to the right SP case, we examine the relations between the right and left scattering coefficients. Then we propose an auxiliary kernel of the left GLME, which allows one to solve the right SP. Using the left TIB algorithm, one can use the relations between the scattering coefficients or the auxiliary kernel to solve the right inverse SP.

The paper is organized as follows. In the introductory section \ref{s:2} we formulate the right and left SPs for the ZS system. The central theoretical section \ref{s:3} devoted to deriving the general relations between the right and left scattering coefficients. 
Here, along with the known results, we introduce a relation between the left and right norming coefficients for the N -soliton solution. In the applied section \ref{s:4} we propose the auxiliary kernel of the left GLME. Finally, in the numerical section \ref{s:5} we show test runs of the TIB algorithm, which illustrate the obtained results by reconstructing the various nonsymmetric potentials from their right scattering data. We finish with the conclusions in section \ref{s:6}. In Appendix (section \ref{s:7}) we present a brief derivation of the connection between the right and left normaing coefficients for the N-soliton solution. 

\section{Statements of right and left scattering problems for the ZS system.}\label{s:2}
The IST method interprets the solution to the NLS equation $u(x)$ as a scattering potential for the spectral wave function $\Phi(x)$ obeying the ZS system (\ref{ZS}). Here we consider two equivalent formulations of SPs for the ZS system in the case of rapidly decaying at infinity $u(x)$: 1) left one when the incident spectral wave function arrives at a right side of the potential 2) the right one when the incident spectral wave function arrives at a left side of the potential. The following asymptotics conditions for $\Phi(x)$ define the right SP:\\
\begin{equation}\label{BCR}
\Phi(x\rightarrow-\infty)\rightarrow
\left(
\begin{array}{c}
e^{-\I\la x} \\
0 \end{array}
 \right);
\qquad
 \Phi(x\rightarrow+\infty)\rightarrow
\left(
\begin{array}{c}
a_{\mr}(\la)e^{-\I\la x} \\
b_{\mr}(\la)e^{\I\la x} \end{array}
 \right),
\end{equation}
where $a_{\mr}(\lambda)$ and $b_{\mr}(\lambda)$ are the scattering coefficients, which are in one to one correspondence with the scattering potential $u(x)$. The index $_\mr$ here indicates the right-hand formulation of the problem. The physical interpretation of the right SP is the following. The incident wave function $(a_\mr(\la)e^{-\I\la x},0)^\mathrm{T}$ propagate from the right side of the potential to the left. The potential partially reflects it, so that the reflected wave function $(0, b_\mr(\la)e^{\I\la x})^\mathrm{T}$ propagates back to the right. The asymptotic at $x\rightarrow-\infty$ in (\ref{BCR}) $(e^{-\I\la x},0)^\mathrm{T}$ represents the normalized to unity transmitted wave function propagating to the left.

In the focusing case ($\sigma=-1$), the eigenvalue spectrum of the ZS system represents the whole real axis (continuous spectrum part) and discrete complex points (discrete spectrum part) $\la_k = \xi_k +\I \eta_k,\;\;k=1,...,M$, where $M$ is the number of discrete components of the spectrum. In contrast, in the defocusing case ($\sigma=-1$) it has only the continuous part occupating the real axis.

The continuous spectrum is characterized by the reflection coefficient $r_\mr$ (spectral reflectance):
\begin{equation*}
r_{\mr} (\xi)=\frac{b_{\mr} (\xi)}{a_{\mr} (\xi)},\;\;\xi \in {\cal R}.
\end{equation*}
Its Fourier transform $R_\mr(z)=\frac{1}{2\pi} \int_{-\infty}^{+\infty}r_\mr(\xi) e^{+\I \xi z} d\xi$ is the pulse response of the system.

In the focusing case, the discrete spectrum points $\la_k$ are defined as roots of the scattering coefficient $a_{\mr} (\la)$:
\begin{equation}\label{Roots}
a_\mr (\la_k )=0, \qquad \eta_k>0.
\end{equation}
Each root of equation (\ref{Roots}) corresponds to a soliton in the wave field $u(x)$.
In addition, the discrete spectrum is characterized by the norming coefficients:
\begin{equation}\label{Rhor}
\rho_{\mr,k}=\frac{b_{\mr} (\la)}{a_{\mr} '(\la)}\biggr|_{\la=\la_k},
\end{equation}
where the prime stands for the derivative with respect to $\la$.

Note, that the scattering coefficient $a_\mr (\la)$ is an analytic function in the upper complex half-plane $\la$, including the real axis, while $a^*_\mr (\la^*)$ is an analytic function in the low complex half-plane \cite{Zakharov:1984,Lamb:1980,Ablowitz:1981,Faddeev:2007}. The scattering coefficients $b_\mr (\la)$ and $b^*_\mr (\la^*)$ have the same analytical properties as $a_\mr (\la)$ and $a^*_\mr (\la^*)$ only in the case when the potential has a compact support (i.e. nonzero in a finite region of space). Otherwise, they do not have an analytical continuation off the real axes, although they are always defined at the points $\la_k$ and $\la^*_k$. This means, that the scattering problem (\ref{BCR}) is defined in general only for $\la=\xi\in {\cal R}$ and $\la = \la_k$.
Constructing the IST scheme, one needs to consider only the real axes and the upper half of the $\la$-plane \cite{Zakharov:1984,Lamb:1980,Ablowitz:1981,Faddeev:2007}.

In the focusing case, the direct SP consists in finding for a given $u(x)$, the full set of the spectral data, which is $\{ a_{\mr} (\la),\, b_{\mr}(\la),\,\la_k,\,\rho_{{\mr},k} \}$ in the focusing case and
$\{ a_{\mr} (\xi),\, b_{\mr}(\xi) \}$ in the defocusing case. The spectral data enters the GLME by forming the kernel of the integral equation $\Omega_{\mr}(z)$, see section \ref{s:3} below.

Numerical methods for solving the direct SP are relatively well developed by now; see the references in \cite{Turitsyn:2017,Mullyadzhanov:2021,Delitsyn:2022}.
In general, the inverse SP reconstructs $u(x)$ by the available set of the spectral data. In this work, we solve the inverse SP using the GLME and assume that the kernel $\Omega_{\mr}(z)$ is a given function.

In the case of the left scattering problem, we use the following asymptotics conditions for $\Phi(x)$:
\begin{equation}\label{BCL}
\Phi(x\rightarrow-\infty)\rightarrow
\left(
\begin{array}{c}
b_{\ml}(\la)e^{-\I\la x} \\
a_{\ml}(\la)e^{+\I\la x} \end{array}
 \right); 
 \qquad
 \Phi(x\rightarrow+\infty)\rightarrow
 \left(
\begin{array}{c}
0 \\
e^{+\I\la x} \end{array}
 \right),
\end{equation}
where the subscript ${\ml}$ indicates the left formulation of the SP. Physically, the boundary conditions (\ref{BCL}) mean that the wave $(0, a_{\ml}(\la)e^{+\I\la x})^\mathrm{T}$ incident to the potential from the left side, reflects as $(b_{\ml}(\la)e^{-\I\la x},0)^\mathrm{T}$ and transmit as $(0, e^{+\I\la x})^\mathrm{T}$ (compare with the physical interpretation of the right SP given above).

The left spectral data: the reflection coefficient, pulse response, discrete spectrum, and the normalization coefficients - are defined similarly to the right SP case:
\begin{equation}
r_{\ml} (\xi)=\frac{b_{\ml} (\xi)}{a_{\ml} (\xi)}, \quad R_{\ml} (z)=\frac{1}{2\pi}\int_{-\infty}^{+\infty}r_{\ml} (\xi) e^{-\I \xi z}d\xi, \qquad \xi \in {\cal R}.
\end{equation}
\begin{equation}
\{\la_k \,|\, a_{\ml}(\la_k)=0, \quad \eta_k>0 \}, \qquad \rho_{\ml,k}=\frac{b_{\ml} (\la)}{a_{\ml}'(\la)}\biggr|_{\la=\la_k}.
\end{equation}

Note, that the roots of $a_{\ml}(\la)$ coincide with the roots of $a_{\mr}(\la)$.
In addition, the left scattering coefficients have the same analytical properties as the right ones, and
the direct and inverse left SPs statements are similar to those for the right SPs case.

The complete IST method scheme also requires equations for the time evolution of the spectral data. These equations have the following form:
\begin{equation}\label {GGKMR1}
	a_{\mr} (\la,t)=a_{\mr} (\la,0),\qquad b_{\mr} (\la,t)=b_{\mr} (\la,0)e^{4\I\la^2 t},	
\end{equation}
\begin{equation}\label {GGKMR2}
	r_{\mr} (\xi,t)=r_{\mr} (\xi,0) e^{4\I\xi^2 t},\qquad \rho_{\mr,k}(t) = \rho_{\mr,k}(0) e^{4\I\la_k^2 t}.		
\end{equation}
for the right SP, while for the left SP case they are written as follows:
\begin{equation}\label {GGKML1}
a_{\ml} (\la,t)=a_{\ml} (\la,0),\qquad b_{\ml} (\la,t)=b_{\ml} (\la,0)e^{-4\I\la^2 t},
\end{equation}
\begin{equation}\label {GGKML2}
r_{\ml} (\xi,t)=r_{\ml} (\xi,0) e^{-4\I\xi^2 t},\qquad \rho_{\ml,k}(t) = \rho_{\ml,k}(0) e^{-4\I\la_k^2 t}.
\end{equation}
Together with algorithms for solving the direct and the inverse SPs, these evolutionary equations allow one to solve the Cauchy problem for the NLSE with a wave field $u(x)$ given at $t=0$.

\section{On the connection between the right and left scattering problems }\label{s:3}
This section derives general relations between the right and left spectral problems and their scattering data.
Our derivation uses the well-known involution property of the ZS system and the properties of its fundamental solutions -- the Jost basis functions, see \cite{Zakharov:1971,Lamb:1980,Ablowitz:1981}, which we briefly remind here.
The involution property of the ZS system (\ref{ZS}) means that the matrix $\widehat{\mathbf{U}}$ satisfy the relation:
\begin{equation}\label{ZSinvolution}
\widehat{\mathbf{U}}^\dag(-\sigma \la^*) = \sigma \widehat{\mathbf{U}}(\la),
\end{equation}
where $^\dag$ means Hermitian conjugate. 
The involution property allows one to construct a new solution (called involution) to the ZS system (\ref{ZS}) from a known one. We write the latter in the following form:
\begin{equation}\label{Phi0}
\Phi_0=
\left(
\begin{array}{c}
\alpha_1 (x,\la) \\
\alpha_2 (x,\la) \end{array}
 \right).
\end{equation}
Substituting (\ref{Phi0}) to the ZS system (\ref{ZS}) we obtain that,
\begin{equation}\label{ZSalpha}
\biggl\{
\begin{array}{c}
(\alpha_1)'_x =-\I\la\alpha_1+u \alpha_2 \\
(\alpha_2)'_x =\sigma u^* \alpha_1+\I\la \alpha_2 
\end{array}.
\end{equation}
After complex conjugation, the system (\ref{ZSalpha}) reads as:
\begin{equation}\label{conjZSalpha}
\biggl\{
\begin{array}{c}
(\alpha_1^*)'_x =\I\la^*\alpha_1^*+u^* \alpha_2^* \\
(\alpha_2^*)'_x =\sigma u \alpha_1^*-\I\la^* \alpha_2^* 
\end{array}.
\end{equation}
Then we multiply the second equation in (\ref{conjZSalpha}) by $\sigma$ and write the resulting system in matrix form:
\begin{equation}
\left(
\begin{array}{c}
\alpha_1^*\\
\sigma\alpha_2^* 
\end{array}
\right)'_x=
\left[
\begin{array}{cc}
\I\la^* & \sigma u^* \\
u & -\I\la^*  \end{array}
 \right]
 \left(
\begin{array}{c}
\alpha_1^*\\
\sigma\alpha_2^* 
\end{array}
\right).
 \end{equation}
Now we rewrite the final system of equations in the following form:
\begin{equation}\label{finalZSalpha}
\left(
\begin{array}{c}
\sigma\alpha_2^*(x,\la)\\
\alpha_1^*(x,\la) 
\end{array}
\right)'_x=
\left[
\begin{array}{cc}
-\I\la^* & u(x)  \\
\sigma u^*(x) &  \I\la^*
\end{array}
 \right]
 \left(
\begin{array}{c}
\sigma\alpha_2^*(x,\la) \\
\alpha_1^*(x,\la)
\end{array}
\right).
 \end{equation}
Thus we have obtained a solution of the ZS system (\ref{ZS}) with the spectral parameter $\la^*$.
After the replacement of $\la$ by $\la^*$ in (\ref{finalZSalpha}) we transform the matrix of the system (\ref{finalZSalpha}) into the matrix $\widehat{\mathbf{U}}$, which means that we arrive at the following new, i.e. linearly independent from (\ref{Phi0}), solution (involution) of the ZS system:
\begin{equation}\label{Phi0Tilde}
\widetilde{\Phi}_0=
\left(\begin{array}{c}
\sigma\alpha_2^*(x,\la^*) \\
\alpha_1^*(x,\la^*)
\end{array}
\right).
\end{equation}
Note that after the transformations made with the ZS system, we obtained again the matrix $\widehat{\mathbf{U}}$ in (\ref{finalZSalpha}) precisely due to the property (\ref{ZSinvolution}).

Now we consider the Jost functions $\Phi_1,\,\Phi_2$ and $\Psi_1,\,\Psi_2$, which are solutions to the ZS system satisfying the following boundary conditions:
\begin{eqnarray}
\Phi_1(x\rightarrow-\infty)\rightarrow
\left(
\begin{array}{c}
e^{-\I\la x} \\
0 \end{array}
 \right), \;\;
 \Phi_2(x\rightarrow-\infty)\rightarrow
 \left(
\begin{array}{c}
0 \\
e^{+\I\la x} \end{array}
 \right),
 \\ \nonumber
\Psi_1(x\rightarrow+\infty)\rightarrow
\left(
\begin{array}{c}
e^{-\I\la x} \\
0 \end{array}
 \right), \;\;
 \Psi_2(x\rightarrow+\infty)\rightarrow
 \left(
\begin{array}{c}
0 \\
e^{+\I\la x} \end{array}
 \right).
\end{eqnarray}

The apparatus of Jost functions has broad applicability in the IST method. The representation of these functions in triangular form allows one to study the analytic properties of the coefficients $a_{\mr} (\la)$, $b_{\mr} (\la)$, $a_{\ml} (\la)$, $b_{\ml} (\la)$ and also to derive the GLME, which, in essence, are the equations for the kernels integrals of this triangular form \cite{Zakharov:1971,Lamb:1980,Ablowitz:1981}.

The pairs $\Phi_1,\,\Phi_2$ and $\Psi_1,\,\Psi_2$ form two linearly independent bases of the solution space of the ZS system.
One can go from one basis to another using a linear transformation with the two-dimensional transition matrixes $\widehat{\mathbf{T}_{\mr}}$ and $\widehat{\mathbf{T}_{\ml}}$ \cite{Zakharov:1971,Lamb:1980,Ablowitz:1981}.

Let us remind ourselves how to derive the transition matrixes.
Based on the definition of the right SP (\ref{BCR}), we have the following equality:
\begin{equation}\label{PhiPsi9}
 \Phi_1 (x,\la)=a_{\mr} (\la) \Psi_1 (x,\la)+b_{\mr} (\la) \Psi_2 (x,\la).
\end{equation}
Applying involution (\ref{Phi0Tilde}) to (\ref{PhiPsi9}) we derive that,
\begin{equation}\label{PhiPsi10}
 \Phi_2 (x,\la^*)=a_{\mr}^* (\la) \Psi_2 (x,\la^*)+\sigma b_{\mr}^* (\la) \Psi_1 (x,\la^*).
\end{equation}
Replacing the spectral parameter $\la$ in (\ref{PhiPsi9}) by its complex conjugate $\la^*$, we obtain:
\begin{equation}\label{PhiPsi11}
 \Phi_2 (x,\la)=a_{\mr}^* (\la^*) \Psi_2 (x,\la)+\sigma b_{\mr}^* (\la^*) \Psi_1 (x,\la).
\end{equation}
Now, from (\ref{PhiPsi9}) and (\ref{PhiPsi11}) we find the right transition matrix as follows:
\begin{equation}\label{PhiPsi12}
\left(
\begin{array}{c}
\Phi_1\\
\Phi_2 
\end{array}
\right)=
\widehat{\mathbf{T}}_{\mr} 
 \left(
\begin{array}{c}
\Psi_1 \\
\Psi_2
\end{array}
\right),
\quad
\widehat{\mathbf{T}}_{\mr}=
\left[
\begin{array}{cc}
a_{\mr}(\la) & b_{\mr}(\la)  \\
\sigma b_{\mr}^*(\la^*) & a_{\mr}^*(\la^*) 
\end{array}
 \right],
 \end{equation}
Note, that the determinant of the right transition matrix $ \det \widehat{\mathbf{T}}_{\mr} = 1$, see \cite{Lamb:1980}.

Similarly, from the definition of the left SP (\ref{BCL}), we derive that,
\begin{equation}\label{PhiPsi13}
 \Psi_2 (x,\la)=b_L (\la) \Phi_1 (x,\la)+a_L (\la) \Phi_2 (x,\la).
\end{equation}
Then, applying the involution (\ref{Phi0Tilde}) to (\ref{PhiPsi13}) we obtain:
\begin{equation}\label{PhiPsi14}
 \sigma \Psi_1 (x,\la)=b_L^* (\la^*) \Phi_2 (x,\la)+\sigma a_L^* (\la^*) \Phi_1 (x,\la).
\end{equation}
Now we multiply equation (\ref{PhiPsi14}) by $\sigma$ (note that $\sigma^2=1$) and find the left transition matrix by writing the equations (\ref{PhiPsi13}--\ref{PhiPsi14}) in matrix form:
\begin{equation}\label{PhiPsi15}
\left(
\begin{array}{c}
\Psi_1\\
\Psi_2 
\end{array}
\right)= \widehat{\mathbf{T}}_{\ml}
 \left(
\begin{array}{c}
\Phi_1 \\
\Phi_2
\end{array}
\right),
\quad
\widehat{\mathbf{T}}_{\ml}=\left[
\begin{array}{cc}
a_{\ml}^*(\la^*) & \sigma b_{\ml}^*(\la^*)  \\
b_{\ml}(\la) & a_{\ml}(\la) 
\end{array}
 \right].
 \end{equation}
 Note, that the determinant of the left transition matrix $ \det \widehat{\mathbf{T}}_{\ml} = 1$, see \cite{Lamb:1980}.
Comparing (\ref{PhiPsi11}) and (\ref{PhiPsi15}) we find that $\widehat{\mathbf{T}}_{\mr} = \widehat{\mathbf{T}}_{\ml}^{-1}$.
 Then, taking into account the unit determinant of  that $\widehat{\mathbf{T}}_{\ml}$ we obtain,
\begin{equation}\label{PhiPsi16}
\left[
\begin{array}{cc}
a_{\mr}(\la) & b_{\mr}(\la)  \\
\sigma b_{\mr}^*(\la^*) & a_{\mr}^*(\la^*) 
\end{array}
 \right]
=\left[
\begin{array}{cc}
a_{\ml}(\la) & -\sigma b_{\ml}^*(\la^*)  \\
-b_{\ml}(\la) & a_{\ml}^*(\la^*) 
\end{array}
 \right].
 \end{equation}
The equality of matrices (\ref{PhiPsi16}) straightforwardly leads to the relations connecting right and left scattering coefficients:
\begin{eqnarray}
\label{PhiPsi17}
a_{\mr}(\la) &=& a_{\ml}(\la),
\\
\label{PhiPsi18}
b_{\mr}(\la) &=& -\sigma b_{\ml}^*(\la^*).
\end{eqnarray}

The relations (\ref{PhiPsi17}--\ref{PhiPsi18}) are valid for any rapidly decaying potentials, regardless of their shapes. 

In the focusing case $(\sigma=-1)$, if $b_R(\la)$ and $b_L(\la)$ are analytical functions in the upper complex plane, we obtain the following formulas for the right and left norming constants of the discrete (soliton) spectrum from(\ref{PhiPsi17}--\ref{PhiPsi18}):
\begin{equation}\label{PhiPsi19}
\rho_{\mr,k}=\frac{b_{\mr} (\la)}{a_{\mr}'(\la)}\biggr|_{\la=\la_k} =\frac{b_{\ml}^*(\la^*)}{a_{\ml}'(\la)}\biggr|_{\la=\la_k}=\frac{b_{\ml}^* (\la^*)}{b_{\ml}(\la)}\biggr|_{\la=\la_k}\rho_{\ml,k}
\end{equation}
In the case of N-soliton potential, $b_R(\la)$ and $b_L(\la)$ have no analytical continuation and equal to zero for every real $\la$. However, the relations between  $\rho_{\mr,k}$ and $\rho_{\ml,k}$ may still be found (see Appendix):
\begin{equation}\label{PhiPsi20}
\rho_{\mr,k}\rho_{\ml,k}=((\la_k-\la^*_k))^2 \left(\prod_{i=1,i \ne k}^N \frac{\la_k-\la^*_i}{\la_k-\la_i}\right)^2.  
\end{equation}
All formulas ( (\ref{PhiPsi17}--\ref{PhiPsi20})) allow one to reduce the right SP to the left one and vice versa. 

Note that according to (\ref{PhiPsi17}), the discrete eigenvalues of the right and left SPs coincide. Also, from (\ref{PhiPsi17}) and (\ref{PhiPsi18}), one can see that the modulus of the right and left reflection coefficients are the same. Thus only the phases of the reflection coefficients and the norming constants are different.

To illustrate the obtained relations, we consider an example of an exact solution to the focusing ZS system in the general case when both continuous and discrete spectra can be present. As such an example, we choose rectangular potential with amplitude $A$ and width $L$, shifted by the value $y$ (otherwise solutions to the right and left SPs coincide) from the origin of $x$ coordinate:
\begin{equation}\label{potential}
u(x,t=0)=\biggl\{
\begin{array}{c}
A,\;\; |x-y|<L/2 \\
0,\;\; |x-y|\ge L/2 
\end{array}
\end{equation}
The direct SP in the case of rectangular potential was first solved in \cite{Manakov:1974}. The right and left scattering coefficients for the potential (\ref{potential}) have the following form:
\begin{equation}\label{ar_rectangle}
a_{\mr} (\la)=e^{\I\la L} \biggl(\cos{(\sqrt{A^2+\la^2} L)} - \frac{\I \la}{\sqrt{A^2+\la^2}} \sin{(\sqrt{A^2+\la^2} L)}\biggr),		
\end{equation}
\begin{equation}\label{br_rectangle}
b_{\mr} (\la)=-\frac{A e^{-2\I\la y}}{\sqrt{A^2+\la^2}} \sin{(\sqrt{A^2+\la^2} L)},		
\end{equation}
\begin{equation}\label{al_rectangle}
a_{\ml} (\la)=e^{\I\la L} \biggl(\cos{(\sqrt{A^2+\la^2} L)} - \frac{\I \la}{\sqrt{A^2+\la^2}} \sin{(\sqrt{A^2+\la^2} L)}\biggr),		
\end{equation}
\begin{equation}\label{bl_rectangle}
b_{\ml} (\la)=-\frac{A e^{+2\I\la y}}{\sqrt{A^2+\la^2}} \sin{(\sqrt{A^2+\la^2} L)},		
\end{equation}
One can verify the relations (\ref{PhiPsi17},\ref{PhiPsi18}) comparing the right scattering coefficients (\ref{ar_rectangle}--\ref{br_rectangle}) with the left ones (\ref{al_rectangle},\ref{bl_rectangle}).

\section{TIB algorithm and GLM equations}\label{s:4}
The Toeplitz inner bordering (TIB) algorithm for the inverse SP for the ZS system was proposed in \cite{Belay:2007} and later improved in \cite {Frumin:2015}.
In addition, the work \cite{Frumin:2015} reports an inverted version of the TIB applicable to the direct SP. The TIB algorithm efficiently solves the discretized integral Gelfand--Levitan--Marchenko equations (GLME) within the left SP framework. See all details on its algorithmic realization in \cite{Belay:2007,Frumin:2015}. In this section, we modify the TIB, allowing one to use it for the right inverse SP.

The GLME for rapidly decaying potentials in the framework of the ZS system has been established in \cite {Zakharov:1971}. We write them in the notations close to monograph \cite{Lamb:1980}. For the left SP at $t=0$, they have the following form:
\begin{equation}\label{GLML1}
A_1^* (x,y)+\int_{-\infty}^x A_2 (x,z) \Omega_{\ml} (y+z)dz = 0,
\end{equation}
\begin{equation}\label{GLML2}
\sigma A_2^* (x,y)+\Omega_{\ml}(x+y)+ \int_{-\infty}^x A_1 (x,z) \Omega_{\ml}(y+z)dz = 0,
\end{equation}
where $A_1 (x,y)$ and $A_2 (x,y)$ are unknown functions to find, while $\Omega_\ml (z)$ is the kernel which is constructed from the left scattering data:
\begin{equation}
\label{LKernel}
\Omega_\ml (z)=\frac{1}{2\pi}\int_{-\infty}^{+\infty} r_\ml(\xi) e^{-\I\xi z} d\xi -\I \sum_k \rho_{\ml,k}e^{-\I \la_k z}.
\end{equation}
The solution to the inverse SP, i.e., the potential $u(x)$, is connected to the solution of the GLME by the following synthesizing relation:
\begin{equation}\label{SynthL}
			u (x)=2\sigma A_2^* (x,x-0) .			
\end{equation}
Similarly, for the right SP, the GLME read as:
\begin{equation}\label{GLMR1}
B_2^* (x,y)+\int_x^{\infty} B_1 (x,z) \Omega_\mr (y+z)dz = 0,
\end{equation}
\begin{equation}\label{GLMR2}
\sigma B_1^* (x,y)+\Omega_\mr(x+y)+ \int_x^{\infty} B_2 (x,z) \Omega_\mr (y+z)dz = 0.
\end{equation}
where $B_1 (x,y)$ and $B_2 (x,y)$ are the unknown functions to find and $\Omega_\mr (z)$ is the kernel, which is constructed from the right scattering data:
\begin{equation}
\label{RKernel}
\Omega_\mr (z)=\frac{1}{2\pi}\int_{-\infty}^{+\infty} r_\mr(\xi) e^{\I\xi z} d\xi -\I \sum_k \rho_{\mr,k}e^{\I \la_k z}.
\end{equation}
The synthesizing relation for the right inverse SP is:
\begin{equation}\label{SynthR}
			u (x)=2\sigma B_1 (x,x+0) .			
\end{equation}
Note that the time dependence for the GLME can be recovered using the evolution equations (\ref{GGKMR1}-\ref{GGKML2}).

The right and left GLME are connected via nontrivial relations between the right and left scattering data derived in the previous section. In addition the formulas for potential (\ref{SynthL}) and (\ref{SynthR}) are different. Nevertheless, we find an easy way to represent the solution of the right GLME via an auxiliary solution of the left GLME (and vice versa), which allows one to solve the right inverse SP using the left GLME (and vice versa). Let us perform the following transformation in the GLME for the right SP:
\begin{equation}\label{ChangeVar}
A_1 (x,y)=B_2^* (-x,-y),\;\;A_2 (x,y)=B_1^*(-x,-y),\;\;\Omega_\ml (x)=\Omega_\mr^* (-x).
\end{equation}
Then we make the change of variables $\{x,y,z\}\rightarrow \{-x,-y,-z\}$ in equations (\ref{GLMR1}), (\ref{GLMR2}), (\ref{SynthR}).
Finally, after complex conjugations of equations (\ref{GLMR1}), (\ref{GLMR2}) we arrive at the system identical to the GLME equations for the left SP problem, i.e., (\ref{GLML1},\ref{GLML2},\ref{SynthL}).
Since the equations coincide, one can use the same TIB algorithm for solving the right and left inverse SPs.

We call the left inverse SP auxiliary and show how to solve the right SP.
The auxiliary inverse SP means solving the left GLME (\ref{GLMR1},\ref{GLMR2}) for an auxiliary kernel $\Omega_{\mathrm{aux},\ml}$ by the standard TIB algorithm \cite{Belay:2007,Frumin:2015}.
The solution of the auxiliary inverse SP is the auxiliary potential $u_\mathrm{aux}(x)$.
Using these notations, we present the following modifications allowing one to solve the right inverse SP by the left TIB algorithm:
\begin{enumerate}

\item Computing the kernel for the auxiliary left inverse SP using (\ref{ChangeVar}):
\begin{equation}\label{InvOmega}
			\Omega_{\mathrm{aux},\ml} (z)=\Omega_\mr^* (-z),	 			
\end{equation}
which for the algorithm means permutation (inversion) of the spatial variable index and complex conjugation of the original right kernel.

\item 	Solving the auxiliary left inverse SP with the kernel (\ref{InvOmega}) by standard TIB algorithm \cite{Belay:2007,Frumin:2015} and finding the auxiliary potential $u_\mathrm{aux}(x)$.

\item 	Obtaining the desired solution to the inverse right SP as follows:
\begin{equation}\label{InvU}
			u(x)=u_\mathrm{aux} (-x),
\end{equation}
which for the algorithm means permutation (inversion) of the spatial variable index in the auxiliary potential.

\end{enumerate}

\section{Numerical examples}\label{s:5}
To illustrate the TIB algorithm's proposed modification, we numerically solve the inverse SP for a couple of potentials with analytically known scattering data.
Choosing the potentials asymmetric concerning the coordinate origin, we guaranty that the right and left direct scattering problems have different solutions, i.e., we present a proper verification of the obtained results.
We start from the truncated hyperbolic secant potential:
\begin{equation}\label{CutOffU}
u (x)=-2A\theta(x)\mbox{sech}(2A x),
\end{equation}
where $A$ represents the potential amplitude, and $\theta(x)$ is the Heaviside step function.

The scattering data of the potential (\ref{CutOffU}) in general contains the continuous and discrete parts, and its right kernel (\ref{RKernel}) can be written as, see \cite{Lamb:1980},
\begin{equation}\label{OmegaCutOff}
\Omega_\mr (z)=2 A \theta(z)\exp(-A z),
\end{equation}

Fig.1 represents a comparison between the exact potential (\ref{RKernel}) and its reconstructed version.
The reconstruction was performed by the modified TIB algorithm using $M=1024$ discretization points on the numerical interval $x\in [0,2]$.
The maximum amplitude of the relative error of the potential reconstruction $\sim 10^{-4}$ and decays
as $M^{-2}$ according to second-order accuracy of the TIB algorithm \cite{Frumin:2015}.
In addition, Fig. 1 shows the kernel $\Omega_\mr (z)$, see Eq. (\ref{OmegaCutOff}), which is defined on the double interval $z\in [0,4]$ according to the GLME.
Note that the TIB algorithm implies that the potential is zero outside the numerical interval so that it has no issues with the discontinuity of the step function in (\ref{OmegaCutOff}).

\vspace*{-7pt}
\begin{figure}[!htb] \centering
\includegraphics[width=0.65\columnwidth]{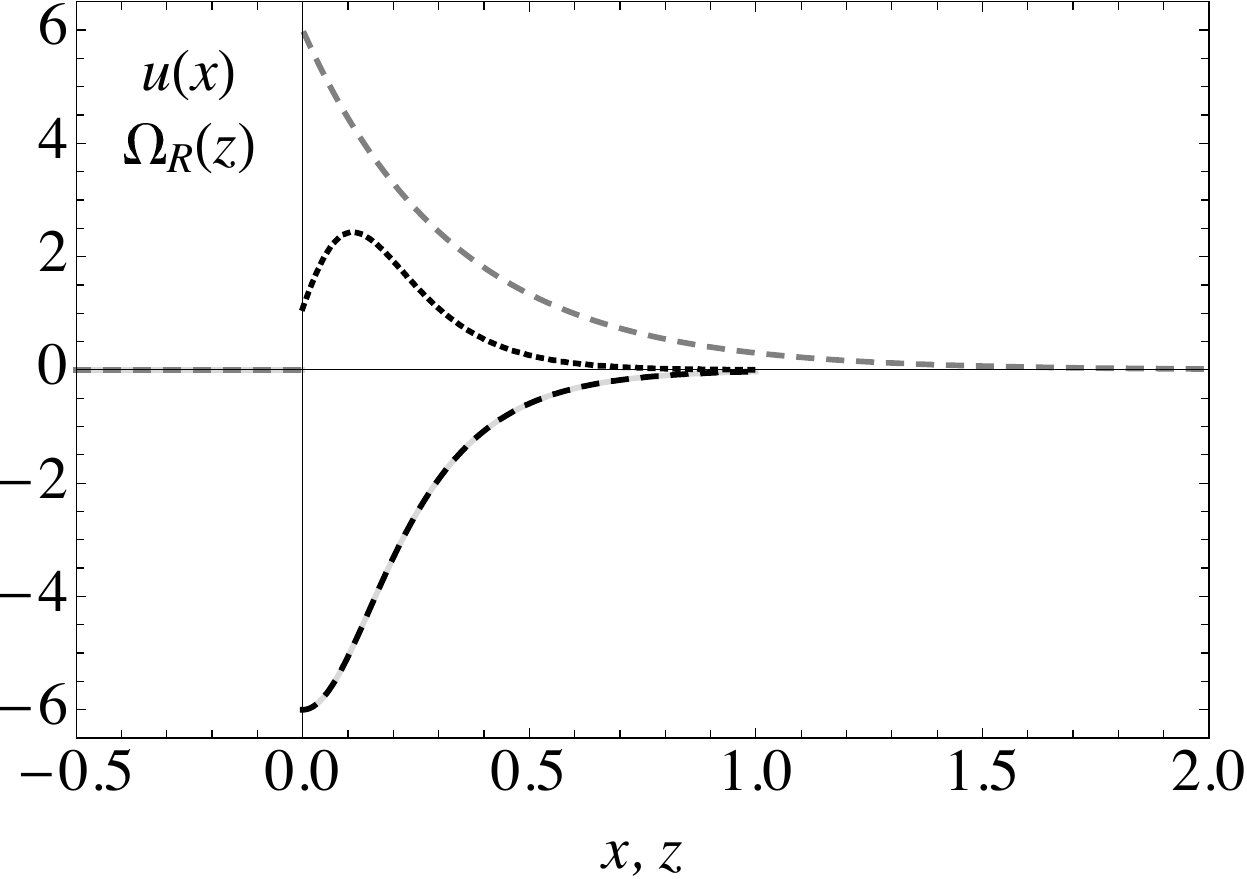}
\caption{Numerical reconstruction of the truncated hyperbolic secant potential (\ref{CutOffU}) with $A=3$ from its right kernel (\ref{OmegaCutOff}) by the use of the modified TIB algorithm.
The solid grey line shows the real part of the exact potential (the imaginary part is zero and not shown here), while the black dashed indicates its reconstructed version.
In addition, the dotted line represents the absolute computational error multiplied by $10^4$, and the grey dashed line shows the right kernel $\Omega_R(z)$ defined on the double interval.}\label{fig:1}
\end{figure}

Next, we consider the full (non-truncated) hyperbolic secant potential at different moments of its time evolution.
At $t=0$ the potential reads as
\begin{equation}\label{sech}
			u(x) = A\;\mbox{sech}(x-x_0 ),
\end{equation}
and its scattering data can be written in the following analytical form, see \cite{Satsuma:1974} and also \cite{Mullyadzhanov:2021}:
\begin{equation}\label{OmegaShift}
			r_\mr (\la)=-\frac{\sin{(\pi A)}}{\mbox{cosh} (\pi \la)} \frac{\Gamma(-\I\la+A+1/2)\Gamma(-\I\la-A+1/2)}{\Gamma^2(-\I\la+1/2)}  e^{-2\I\la x_0 },
\end{equation}
\begin{equation}
			\la_k=\I(A-k+1/2),\;\;k=1,...,\mbox{Integer}[A+1/2],
\end{equation}
\begin{equation}
\rho_{\mr,k}=-\I \frac{\Gamma (2A+1-k) e^{-2\I\la_k x_0}}{\Gamma^2(A+1-k) \Gamma(k)}.
\end{equation}
Here $\Gamma (y)$ is the gamma function of the complex variable $y$. If $A>1/2$, the hyperbolic secant potential contains at least one soliton in its scattering data.
In the general case, the scattering data (\ref{OmegaShift}) contains a nonzero continuous spectrum, except special cases $A=(N-1/2)$, when the reflection coefficient is zero, and the potential represents pure $N$-soliton solution.

Using formulas (\ref{OmegaShift}) and (\ref{GGKMR2}), we find the time evolution of the right kernel of the hyperbolic secant potential as,
\begin{equation}\label{sechOmega}
\Omega_\mr (z,t)=\frac{1}{2\pi}\int_{-\infty}^{+\infty} r_\mr (\xi) e^{\I\xi z + 4 \I \xi^2 t} d\xi -\I 
\sum_k \rho_{\mr,k}(0)e^{\I \la_k z+4 \I \la_k^2 t},
\end{equation}
which allows solving the inverse SP numerically at any time moment, i.e., to find the solution of the Cauchy problem with the initial condition (\ref{sech}).

We choose the potential amplitude $A=1.75$ so that its scattering data contains both discrete and continuous parts.
We also set $x_0=-2$ to make the potential non-symmetric concerning the coordinate origin.
Then we run the modified TIB algorithm at $t=0$ and $t=1.5$ on the interval $[-20,20]$ with the number of discretization points $M=1024$.
At $t=0$ we compare the restored potential with its exact version (\ref{sech}), while for $t=1.5$ we obtain the potential to compare by numerical integration of the NLSE (\ref{NLSE}) using the Split-Step Fourier method (SSFM) \cite{Agrawal:2001,Taha:1984}. See Fig.2. which summarizes these results.

\vspace*{-7pt}
\begin{figure}[!htb] \centering
\includegraphics[width=0.48\columnwidth]{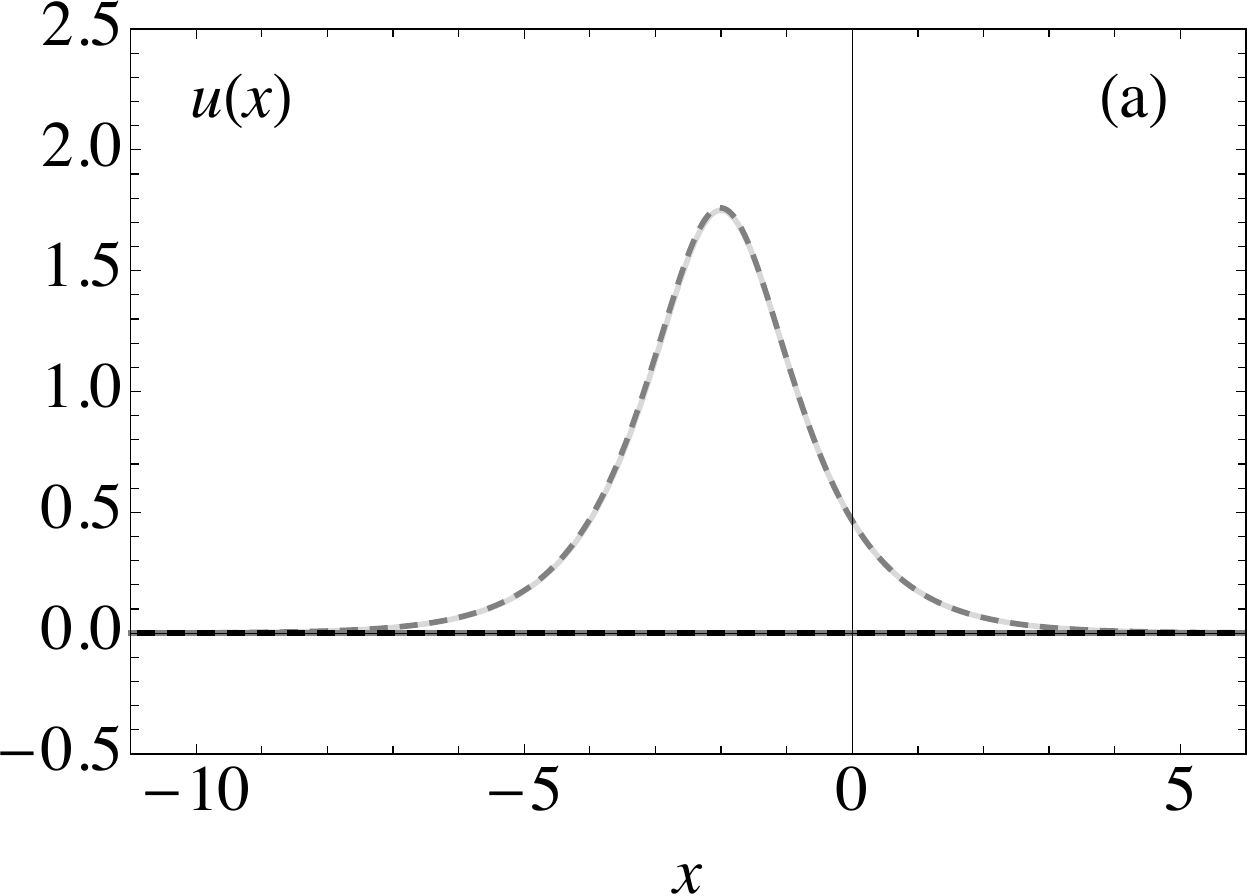}\,\,
\includegraphics[width=0.48\columnwidth]{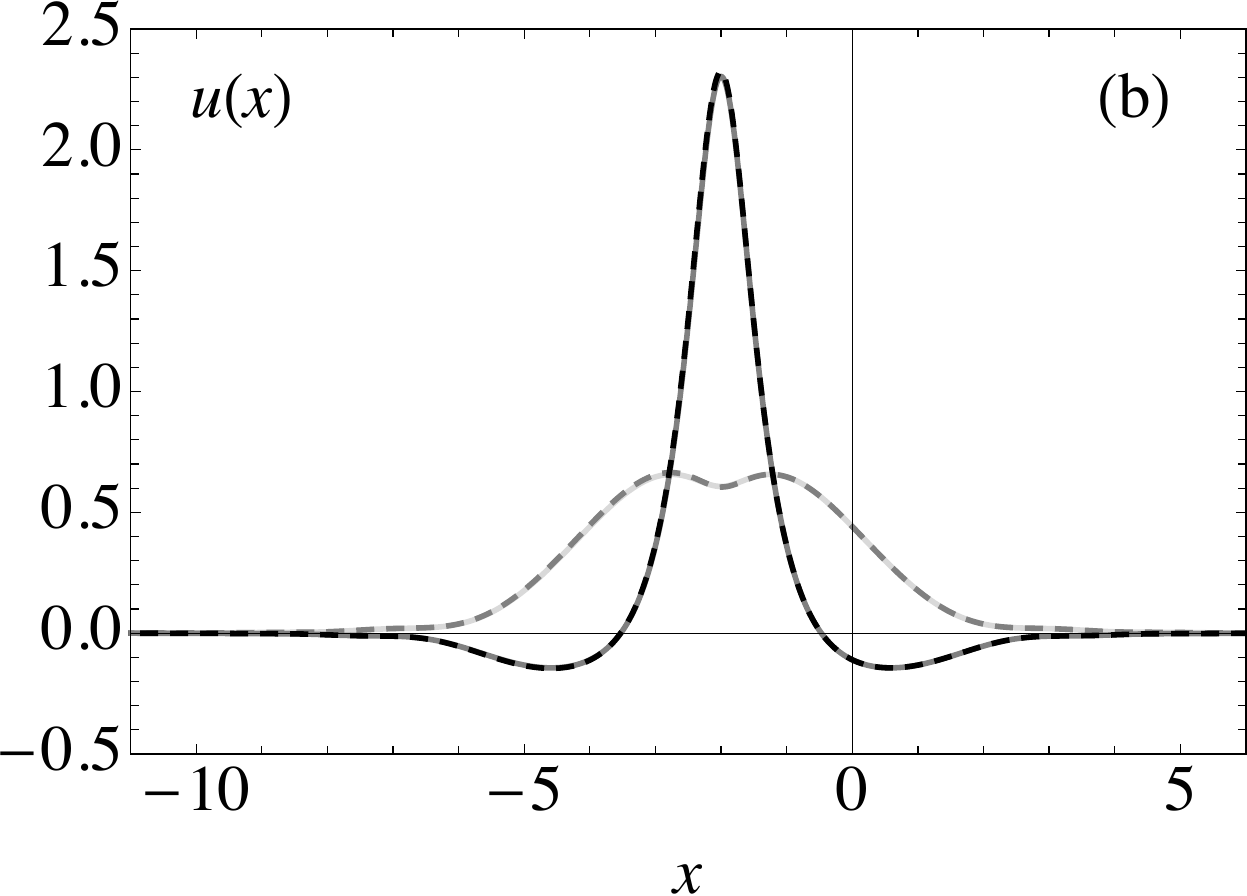}
\caption{Numerical reconstruction of the hyperbolic secant potential (\ref{sech}) with $A=1.75$ and $x_0=-2$ from its right kernel (\ref{sechOmega}) at $t=0$ (a) and $t=1.5$ (b) by the use of the modified TIB algorithm. The black dashed lines, and the gray dashed lines show imaginary and real parts of the restored potential.
The solid grey lines and the light grey solid lines show the same parts of the potential (\ref{sech}) and its numerical evolution computed using SSFM.}
\label{fig:2}
\end{figure}

\section{Conclusion}\label{s:6}
In this work, we have considered right and left formulations of the Zakharov--Shabat scattering problem in light of the numerical solving of the corresponding inverse scattering problem by the TIB algorithm. The general relations between the right and left scattering data described in section \ref{s:3} reveal the nontrivial connection between the two formulations. Along with the known results, we introduce a  relation between the left and right norming coefficients for the N -soliton solution (see the Appendix \ref{s:7} for details).

We have proposed an auxiliary kernel of the left Gelfand--Levitan--Marchenko equations, which allows one to solve the right scattering problem. With the obtained formulas, we have modified the TIB algorithm, previously developed in the left formulation \cite{Belay:2007,Frumin:2015}, to the right scattering problem case.
When the scattering coefficients are known in the $\la$-plane, one can use the relations from section \ref{s:3} to transform the right scattering data into the left formulation and then run the standard TIB algorithm.

In the general case, one can apply the modified TIB algorithm to right scattering data using the auxiliary procedure described in \ref{s:4}.

To verify the proposed modified TIB algorithm, we have applied it for reconstructing truncated and non-truncated asymmetric hyperbolic secant potentials.
We previously applied the TIB algorithm to the potentials with either pure continuous or pure discrete scattering data \cite{Belay:2007,Frumin:2015,Frumin:2017}.
Here we have chosen scattering data containing both discrete and continuous parts, illustrating the applicability of the TIB algorithm in this general case.
In addition, the presented example with the time-evolving kernel shows how the TIB solves the Cauchy problem for the NLSE.

We believe that the presented results will contribute to the field of nonlinear science and applications, providing a universal numerical tool for solving the Zakharov--Shabat inverse scattering problem in different settings. For example, it is used in such rapidly developing fields of studies as nonlinear optical telecommunications \cite{Frumin:2017,Turitsyn:2017} and for numerical design of nonlinear wave field with desired scattering data \cite{Gelash:2018,Suret:2020}. In addition, the obtained results can be generalized to other integrable systems, such as Korteweg-de Vries or Sin-Gordon equations \cite{Zakharov:1984}.

The Zakharov-Shabat system is formally a special case of the AKNS system (see, for example, \cite{Trogdon:2021}). However, the SPs for AKNS is a completely different problem of finding not one, but two unknown potentials. In this case, the data for the left and right scattering problems are often used simultaneously. However, this does not lead to any important consequences for the scattering and norming  coefficients for the Zakharov-Shabat SPs with only one potential.

\acknowledgments{The author is grateful to Professors D.~A. Shapiro and S.~K. Turitsyn for helpful discussions, recommendations, and interest in this work.\\This work supported by the Russian Science Foundation (RSF) (22-22-00653).}

\section{Appendix: Derivation of the relationship between right and left norming coefficients}\label{s:7}
In this Appendix we present the derivation of the relation formula between the right and left norming coefficients $\rho_{\mr,k}$ and $\rho_{\ml,k}$  for the N-soliton solution. To do this, we use the dressing method, following the work of \cite{Gelash:2020}. Usually, the use of the dressing method implies the construction of an exact multisoliton solution of the NLSE (or other integrable equations) by adding one soliton in one dressing step. Here, as in the cited work, we consider the problem of finding the scattering coefficients $a_{tr}(\la)$, $b_{\mr,tr}(\la)$ and $b_{\ml,tr}(\la)$, for trancated N-soliton potentials on a finite interval $[-L;L]$ which are equal to zero outside the interval.

\subsection{Dressing method for the ZS system}
From the point of view of the dressing method, the N-soliton solution $\psi_{NSS}(x)$  is described by its eigenvalues $\la_k=\xi_k+\I\eta_k$, where $k=1,...,N$, and complex numbers $C_k$, which are also called phase factors. The $C_k$ are different from the norming coefficients $\rho_{\mr,k}$ and $\rho_{\ml,k}$ but the connection between them will be found below.

Without loss of generality, we set $t=0$, and also $\eta_1\ge\eta_2\ge...\ge\eta_N$.

Let us introduce $\hat{\Phi}_{(n)} (x,\la)$ as the matrix of the Fundamental System of Solutions (FSS) of the ZS equations for the potential $\psi_n(x)$, where $\psi_n(x)$ is the NLSE multisoliton solution containing solitons from the first to the Nth. The dressing method assumes the recurrence relation:
\begin{equation}\label{A1}
\hat{\Phi}_{(n)} (x,\la)=\hat{{\bf\chi}}_n(x,\la) \hat{{\bf \Phi}}_{(n-1)}(x,\la), 
\end{equation}
where $\hat{\bf\chi}_n(x,\la)$ is called the dressing matrix.
With the known $\hat{{\bf\chi}}_{(n-1)}(x,\la)$ the dressing matrix is described by the following equation:
\begin{eqnarray}\label{A2}
\hat{{\bf\chi}}_n(x,\la)=
\left[
\begin{array}{cc}
1 & 0 \\
0 & 1  \end{array}
\right]
+\frac{\la_n-\la_n^*}{\la-\la_n}\frac{1}{|{\bf q}_n|^2}
\left[
\begin{array}{cc}
q^*_{(n)1}q_{(n)1} & q^*_{(n)1}q_{(n)2} \\ \nonumber
q^*_{(n)2}q_{(n)1} & q^*_{(n)2}q_{(n)2}  \end{array}
\right], \\
{\bf q}_n=\hat{{bf \Phi}}_{(n-1)}(x,\la_n^*)
\left[
\begin{array}{c}
1   \\
C_n  \end{array}
\right]
\end{eqnarray}
The `'zero" matrix FSS can be considered the trivial soliton-free solution of the NLSE, $\psi_0(x)=0$. In this case
\begin{equation}\label{A3}
\hat{{\bf \Phi}}_{(0)}(x,\la)=
\left[
\begin{array}{cc}
e^{-\I\la x} & 0 \\
0 & e^{\I\la x}  \end{array}
\right]
\end{equation}
Formula (\ref{A1}) can be understood as an elementary step of the cyclic algorithm for finding $\hat{{\bf\Phi}}_{(N)}(x,\la)$ -- the N-solton solution (NSS) matrix for $\psi_{NSS}(x)$. Knowing $\hat{{\bf\Phi}}_{(N)}(x,\la)$ we can find the scattering coefficients of the trancated (cutoff) potential $\psi_{NSS,tr}(x)=\psi_{NSS}(x) \theta(|x-L|)$

\subsection{Solution of the ZS system at points $x=\pm L$ for the potential $\psi_{NSS}(x)$}
Following the work \cite{Gelash:2020}, slightly changing notation, we write:

\begin{equation}\label{A4}
\hat{{\bf\Phi}}_{(N)}(L,\la)=
\left[
\begin{array}{cc}
(1+{\it o}_N)e^{-\I\la L} & (-\beta^*_{N,+L}(\la^*)+{\it o}_Ne^{2\I\la L})e^{-\I\la L} \\
 \left(\frac{-\beta_{N,+L}(\la)}{a_N(\la)}+{\it o}_N e^{-2\I\la L}\right)e^{\I\la L} &(a_N^*(\la^*)+{\it o}_N) e^{\I\la L}  \end{array}
\right],
\end{equation}
where
\begin{eqnarray}\label{A5}
\beta_{N,+L}(\la)=a_N(\la)\sum_{i=1}^N \frac{B_{N,i}}{C_i}\frac{e^{-2\I(\la-\la_i)L}}{\la-\la_i};\\ \nonumber
a_N(\la)=\prod_{i=1}^N\frac{\la-\la_i}{\la-\la^*_i};\;\;B_{N,i}=(\la_i-\la_i^*)\prod_{i=1, j\ne i}^N\frac{\la_i-\la_j^*}{\la_i-\la_j}.
\end{eqnarray}
The value $L$ is considered to be much greater than the distances to the soliton peaks from the center of coordinates, $a_N (\la)$ is the scattering coefficient of the N-soliton solution, and ${\it o}_N$ includes small terms of the series expansion (for $L \rightarrow \infty$) and is defined as
\begin{equation}\label{A6}
{\it o}_N={\it o}\left(\exp{\left(-2L  \min_{i,j=1,...,N} |\eta_i-\eta_j|\right) }\right).
\end{equation}

By analogy, we write out the asymptotic of the matrix $\hat{{\bf\Phi}}_{(N)}(x=-L,\la)$:
\begin{equation}\label{A7}
\hat{{\bf\Phi}}_{(N)}(-L,\la)=
\left[
\begin{array}{cc}
(a_N^*(\la^*)+{\it o}_N) e^{\I\la L}  &   \left(\frac{-\beta^*_{N,-L}(\la^*)}{a_N(\la)}+{\it o}_N e^{-2\I\la L}\right)e^{\I\la L}\\
 (\beta_{N,-L}(\la)+{\it o}_Ne^{2\I\la L})e^{-\I\la L}& (1+{\it o}_N)e^{-\I\la L} \end{array}
\right],
\end{equation}
where
\begin{equation}\label{A8}
\beta_{N,-L}(\la)=a^{-1}_N(\la)\sum_{i=1}^N C^*_i B^*_{N,i} \frac{e^{2\I(\la-\la^*_i)L}}{\la-\la^*_i}
\end{equation}
Recall that $a_N^* (\la^* )=a_N^{-1} (\la)$ follows from (\ref{A6}). In work  \cite{Gelash:2020} the value $\beta_{N,-L}(\la)$ is neglected, including it in the remainder term ${\it o}_N$, and the connection between the phase factors and the right norming coefficients is obtained. However, we take this value into account for a more clear consideration.

\subsection{Norming coefficients of the truncated N-soliton potential}
	The solution of the direct right scattering problem for $\psi_{NSS,tr} (x)$ must take into account the right boundary conditions: 
	
$\left[
\begin{array}{c}
\psi_1   \\
\psi_2 \end{array}
\right](-L,\la)=\left[
\begin{array}{c}
e^{+\I\la L}   \\
0  \end{array}
\right] $;\;\;\;\; 
$\left[
\begin{array}{c}
\psi_1   \\
\psi_2 \end{array}
\right](L,\la)=\left[
\begin{array}{c}
a_{tr}(\la)e^{-\I\la L}   \\
b_{tr,+L}(\la)e^{+\I\la L}  \end{array}\right] $. 

This solution satisfies the equation
$\left[
\begin{array}{c}
\psi_1   \\
\psi_2 \end{array}
\right] \cong a_N(\la)(\Psi_{(N)I}-\beta_{N,-L}(\la) \Psi_{(N)II})$, where $\Psi_{(N)I}$ and $\Psi_{(N)II}$ are the first and the second columns of the FSS, respectively. This equality is approximate, since instead of zero at the point $x=-L$, the function $\psi_2$ will take the value ${\it o}_N e^{\I\la L}$. 
In addition, we assume that $\beta_{N,-L}^* (\la^* )\beta_{N,-L} (\la)={\it o}_N$. 

Considering the relation at the point $x=+L$ and using the formula (formula for linking left and right $b$), we get:
\begin{eqnarray}\label{A9}
b_{N,+L}(\la)=\beta_{N,+L}(\la)-\beta_{N,-L}(\la),\\ \nonumber    
b_{N,-L}(\la)=\beta_(N,+L)^* (\la^* )-\beta_{N,-L}^* (\la^* ).    
\end{eqnarray}
A detailed examination shows that $b_{N,+L}(\la)$, as well as $b_{N,-L} (\la)$, have simple poles at points of the discrete spectrum. 
Therefore, the definition $\lambda_k =\frac{b(\la)}{a^* (\la)}|_{\la=\la_k}$ of the norming coefficient is not valid. Instead, it is necessary to look for the norming coefficients $\rho_{\mr,k}$ and $\rho_{\ml,k}$ as residues of the corresponding reflection coefficients at the corresponding points. 
As a result, we get:
\begin{eqnarray}\label{A10}
\rho_{\mr,k}=  \frac{\la_k-\la^*_k}{C_k} \prod_{i=1, i \ne k}^N \frac{\la_k-\la^*_i}{\la_k-\la_i},\\ \nonumber   
\rho_{\ml,k}= C_k(\la_k-\la^*_k)\prod_{i=1,i \ne k}^N \frac{\la_k-\la^*_i}{\la_k-\la_i}.
\end{eqnarray}
The first formula was obtained earlier by the authors of \cite{Gelash:2020}. 
To finally derive the relationship between the coefficients $\rho_{\mr,k}$ and $\rho_{\ml,k}$, it suffices to multiply both expressions (\ref{A10})
\begin{equation}\label{A11}
\rho_{\mr,k} \rho_{\ml,k}=(\la_k-\la^*_k)^2 \left(\prod_{i=1,i \ne k}^N \frac{\la_k-\la^*_i}{\la_k-\la_i}\right)^2.  
\end{equation}

\end{document}